\documentclass[final,5p,times,twocolumn]{elsarticle}
\usepackage{graphicx}
\usepackage{amssymb}
\usepackage{amsmath}
\usepackage{bm}
\usepackage{mathtools, cuted}
\usepackage{color}
\usepackage{hyperref}
\usepackage{verbatim}
\usepackage{verbatimbox}
\usepackage{appendix}
\usepackage{float}

\usepackage{listings}
\usepackage{xcolor}
\usepackage[most]{tcolorbox}
\usepackage{pifont}
\usepackage{upquote}

\usepackage{array}
\usepackage{booktabs}
\usepackage{threeparttable}
\usepackage{multirow}
\usepackage{makecell}
\usepackage{arydshln}
\hypersetup{colorlinks = true,linkcolor = blue,anchorcolor =red,citecolor = blue,filecolor = red,urlcolor = red, pdfauthor=author}

\definecolor{codegreen}{rgb}{0,0.6,0}
\definecolor{codegray}{rgb}{0.5,0.5,0.5}
\definecolor{codepurple}{rgb}{0.58,0,0.82}

\lstdefinelanguage{QE}{
    keywords={CONTROL,SYSTEM,ELECTRONS,ATOMIC_SPECIES,ATOMIC_POSITIONS,K_POINTS,IONS,CELL,INPUTPP,INPUTPH,PLOT,ENERGY_GRID,INPUT},
    morecomment=[l]{\#}
}

\lstdefinelanguage{Command}{
    keywords={cd,ls,mkdir,mv,rm,cp,pw.x,mpirun,ph.x,tail,vi,for,xcrysden,grep},
    morecomment=[l]{\#}
}

\newtcblisting{codepython}[2][]{breakable, enhanced, 
  arc=3pt, top=0mm, bottom=0mm, coltitle=black, colframe=gray!30!white,
  fonttitle=\bfseries, listing only, drop fuzzy shadow=gray!25!white,
  listing options={basicstyle=\footnotesize\ttfamily,language=Python,upquote=true,showstringspaces=false,commentstyle=\color{blue}},
  title=#2}

\newtcblisting{codeQE}[2][]{breakable, enhanced, 
  arc=3pt, top=0mm, bottom=0mm, coltitle=black, colframe=gray!30!white,
  fonttitle=\bfseries, listing only, drop fuzzy shadow=gray!25!white,
  listing options={basicstyle=\footnotesize\ttfamily,language=QE,upquote=true,showstringspaces=false,commentstyle=\color{blue}},
  title=#2}

\newtcblisting{codeQEsmall}[2][]{breakable, enhanced, 
  arc=3pt, top=0mm, bottom=0mm, coltitle=black, colframe=gray!30!white,
  fonttitle=\bfseries, listing only, drop fuzzy shadow=gray!25!white,
  listing options={basicstyle=\scriptsize\ttfamily,language=QE,upquote=true,showstringspaces=false,commentstyle=\color{black}},title=#2}

\newtcblisting{terminal}[2][]{breakable, enhanced, arc=3pt, top=0mm,
  bottom=0mm, left=-7mm, right=-2mm, coltitle=black, colframe=gray!30!white,
  fonttitle=\bfseries, listing only, drop fuzzy shadow=gray!25!white,
  listing options={numbers=none,frame=none,basicstyle=\ttfamily,language=Bash,upquote=true,showstringspaces=false,escapeinside={*}{*}},
  title=#2}

\newtcblisting{terminal2}[2][]{breakable, enhanced, arc=3pt, top=0mm,
  bottom=0mm, left=-7mm, right=-2mm, coltitle=black, colframe=gray!30!white,
  fonttitle=\bfseries, listing only, drop fuzzy shadow=gray!25!white,
  listing options={numbers=none,frame=none,basicstyle=\footnotesize\ttfamily,language=Bash,upquote=true,showstringspaces=false,escapeinside={*}{*}},
  title=#2}
  

\definecolor{codegreen}{rgb}{0,0.6,0}
\definecolor{codegray}{rgb}{0.5,0.5,0.5}
\definecolor{codepurple}{rgb}{0.58,0,0.82}
\definecolor{backcolour}{rgb}{0.95,0.95,0.92}
\journal{Computer Physics Communications}

\begin{document}

\begin{frontmatter}

\title{{\sc QR$^2$-code}: An open-source program for double resonance Raman spectra}
\author[a]{Jianqi Huang\corref{author}}
\author[b,c]{Renhui Liu}
\author[b,c]{Ye Zhang}
\author[d]{Nguyen Tuan Hung\corref{author}}
\author[e]{Huaihong Guo}
\author[f,g]{Riichiro Saito}
\author[b,c]{Teng Yang\corref{author}}

\cortext[author] {Corresponding authors.\\\textit{E-mail address:} jqhuang@lam.ln.cn, nguyen.tuan.hung.e4@tohoku.ac.jp; yanghaiteng@msn.com}
\address[a]{Liaoning Academy of Materials, Shenyang 110167, China}
\address[b]{Shenyang National Laboratory for Materials Science, Institute of Metal Research, Chinese Academy of Sciences, Shenyang 110016, China}
\address[c]{School of Materials Science and Engineering, University of Science and Technology of China, Shenyang 110016, China}
\address[d]{Frontier Research Institute for Interdisciplinary Sciences, Tohoku University, Sendai 980-8578, Japan}
\address[e]{College of Sciences, Liaoning Petrochemical University, Fushun 113001, China}
\address[f]{Department of Physics, National Taiwan Normal University, Taipei 11677, Taiwan}
\address[g]{Department of Physics, Tohoku University, Sendai 980-8578, Japan}

\begin{abstract}
We present an open-source program {\sc QR$^2$-code} that computes double-resonance Raman (DRR) spectra using first-principles calculations. {\sc QR$^2$-code} can calculate not only two-phonon DRR spectra but also single-resonance Raman spectra and defect-induced DRR spectra. For defect-induced DDR spectra, we simply assume that the electron-defect matrix element of elastic scattering is a constant. 
Hands-on tutorials for graphene are given to show how to run {\sc QR$^2$-code} for single-resonance, double-resonance, and defect-induced Raman spectra. We also compare the single-resonance Raman spectra by {\sc QR$^2$-code} with that by {\sc QERaman} code. In {\sc QR$^2$-code}, the energy dispersions of electron and phonon are taken from {\sc Quantum ESPRESSO} ({\sc QE}) code, and the electron-phonon matrix element is obtained from the electron-phonon Wannier ({\sc EPW}) code.  
All codes, examples, and scripts are available on the GitHub repository.
\end{abstract}


\end{frontmatter}
{\bf PROGRAM SUMMARY}

\begin{small}
\noindent
{\em Program Title:} {\sc QR$^2$-code}\\
{\em Developer's repository link:} \href{https://github.com/JoeyyHuang/QR2-code}{https://github.com/JoeyyHuang/QR2-code}\\
{\em Licensing provisions:} GNU General Public Licence 3.0\\
{\em Programming language:} Fortran\\
{\em External routines}: {\sc Quantum ESPRESSO v7.3.1}, {\sc EPW v5.8.1}\\
{\em Nature of problem:}  A DFT-based program for computation and analysis of the single, double resonance, and defect-induced Raman spectra. \\
{\em Solution method:} The Raman spectrum is calculated by the time-dependent perturbation theory, in which the electron-photon and electron-phonon matrix elements are obtained from the modified packages {\sc Quantum ESPRESSO} and {\sc EPW}, respectively.\\
{\em Supplementary material:} \href{http://qr2-code.com}{http://qr2-code.com}
\end{small}

\section{Introduction}\label{sec:intro}
Raman spectra of low-dimensional materials are frequently observed by changing the laser excitation energies, in which the resonance Raman effect is essential not only for assigning the phonon modes but also for the electronic structure of materials ~\cite{jorio2011raman,malard2009raman,i1335}. 
In the previous work, we have developed an open-source program, {\sc QERaman}~\cite{hung2024qeraman}, for calculating single-resonance Raman (SRR) spectra by using the {\sc Quantum ESPRESSO (QE)} code~\cite{giannozzi09-espresso}, an open-source package for first-principles calculations. 
However,  {\sc QERaman} does not support the double-resonance Raman (DRR) spectra, which are often observed in Raman spectra of low-dimensional materials. 
For example, seven peaks in the Raman spectra of the MoTe$_2$ monolayer, which are assigned to DRR peaks, have been observed experimentally from 150 to 350 cm$^{-1}$~\cite{huang2022first}. 
The double-resonance Raman peaks have also been observed on graphene~\cite{zhang2025duv}, h-BN~\cite{saito2024deep}, MoS$_2$~\cite{gontijo2019temperature,liu2024helicity}, graphite~\cite{thomsen2000double}, or single-wall carbon nanotubes~\cite{kurti2002double,saito2003double}. 
Therefore, simultaneously calculating the SRR and DRR spectra is useful to analyze the observed Raman spectra, particularly for low-dimensional materials, since low-dimensional materials often show sharp DRR peaks because of flat phonon dispersion.

The DRR spectra require electron-phonon coupling matrix elements as a function of phonon wavevector $\bm q$ over the entire Brillouin zone (BZ). 
This requires a great deal of computation time compared with the SRR spectra, in which we only need the electron-phonon matrix element at the $\Gamma$ point in the BZ~\cite{hung2024qeraman}. 
In order to avoid this difficulty, several methods have been developed to reduce the computation times of the DRR calculation. 
For graphene or carbon nanotubes, several groups, including the author's group, have developed a tight-binding method for calculating the DRR spectra~\cite{kurti2002double,dresselhaus2005raman,jorio2005resonance}. 
Recently, Huang \textit{et al.}~\cite{huang2022first,liu2024helicity,zhang2022quantum,huang2022new,pang2022accurate} made a new program, named {\sc QR$^2$-code}, for calculating DRR spectra by using {\sc QE} code~\cite{giannozzi09-espresso} combined with the electron-phonon Wannier ({\sc EPW})  code~\cite{ponce2016epw}. 
The {\sc QR$^2$-code} offers three advantages compared to the {\sc QERaman}. 
First, {\sc QR$^2$-code} uses maximally localized Wannier functions in {\sc EPW} to evaluate the electron-phonon matrix element as a function of $\bm q$. 
The EPW calculation thus enables DRR calculations within a reasonably short computational time. It is noted that in {\sc QERaman}, the electron-phonon matrix element is obtained by the modified \verb|ph.x| in {\sc QE}.
Second, {\sc QR$^2$-code} accounts for the Fr\"ohlich electron-phonon coupling, which describes the interaction between electrons and long-wavelength longitudinal optical phonons and is implemented within {\sc EPW}. 
Third, by assuming that the electron-defect matrix element for elastic scattering is constant, the {\sc QR$^2$-code} can produce a defect-induced DRR spectra, though we can not discuss the relative intensity of defect-induced DRR spectra to two-phonon DRR spectra. 
The goal of the paper is to give the open-source of {\sc QR$^2$-code}, which could be useful for the community. We hope that many users will test the program and that they will give any comments to improve {\sc QR$^2$-code}.  

The organization of this paper is as follows. In Sec.~\ref{sec:raman}, we briefly introduce the quantum theory of resonance Raman spectrum. In Sec.~\ref{sec:installandrun}, we explain the installation and usage of {\sc QR$^2$-code}. In Sec.~\ref{sec:example}, we show examples of SRR, two-phonon DRR, and defect-induced DRR of graphene. Other examples for MoS$_2$ monolayer and bulk h-BN are also available in the GitHub repository. We also compare the results of {\sc QR$^2$-code} with {\sc QERaman}. Finally, a summary is given in Sec.~\ref{sec:summary}.

\section{Quantum theory for resonance Raman spectrum}
\label{sec:raman}
A Raman scattering process comprises an incident-photon absorption of an electron with the electron wavevector $\bm{k}$ in the valence band, electron-phonon scattering of the photo-excited electron with one or two phonons (or one elastic-scattering + one phonon), and a scattered-photon emission by recombining the photo-excited electron with a hole at the $\bm{k}$~\cite{f1020,i1179}.
Here, we consider three kinds of resonance Raman processes, i.e., SRR, DRR, and defect-induced DRR processes. Here, resonance means that the intermediate state of the photo-excited electron is a real electronic state, which makes the scattering amplitude become large, which is known as the resonance Raman effect. 
The SSR process involves only one scattering of the photo-excited electron with a phonon with the phonon wavevector $\bm{q}$ near the $\Gamma$ point in the BZ, that is, $\bm{q}=\mathbf{0}$ is required to recombine the scattered photo-excited electron with a hole.
On the other hand, the DRR involves two phonons that possess opposite momentum $\bm{q}$  and $-\bm{q}$  ($\bm{q}\ne \mathbf{0}$), in which the non-zero $\bm{q}$ is selected in the BX with a relatively large electron-phonon matrix element. 
The defect-induced DRR is a special case of DRR, where either the first or the second electron-phonon scattering is replaced by an elastic scattering by a defect of the lattice. A typical example of defect-induced DRR is the D and D$'$ bands in graphene \cite{malard2009raman}. 
Raman spectra plot the intensity of the scattered photon as a function of Raman shift, $\omega_\text{RS}$, in units of cm$^{-1}$.
The detailed derivation of the formula is given in our previous articles~\cite{huang2022first}.

In the following, we will show the Raman scattering amplitude for each Raman process that is given by time-dependent perturbation theory, in which electron-photon and electron-phonon interactions are the perturbation Hamiltonian. 

\subsection{Single-resonance Raman intensity}
\label{sec:I-srr}
Intensity of the SRR scattering as a function of $\omega_\text{RS}$ is expressed by 
\begin{equation}
    I(\omega_\text{RS}) \propto \sum_\mu\left|\mathbf{P}_s^{\dagger}\cdot{\mathbf{R}}(\mu, E_\mathrm{L})\cdot\mathbf{P}_i\right|^2\delta(\omega_\text{RS}\pm\hbar\omega_\mu),
    \label{eq:SRR}
\end{equation}
where $\mathbf{P}_i$ and $\mathbf{P}_s$ are the polarization directions of incident light and scattered light, respectively, and $E_\mathrm{L}$ denotes the energy of incident light. 
The summation on $\mu$ is taken for the $\mu$-th phonon mode at the $\Gamma$ point, and $\omega_\mu$ denotes the phonon frequencies of the $\mu$-th phonon mode. 
The Dirac delta function means that Raman spectra may appear $\omega_\text{RS} \pm\hbar\omega_\mu = 0$. The $+$ ($-$) sign corresponds to the anti-Stokes (Stokes) process. 
${\mathbf{R}}(\mu, E_\mathrm{L})$ is the Raman tensor that expresses the Raman scattering by the $\mu$-th phonon for the laser excitation energy $E_\mathrm{L}$. 
Using a third-order time-dependent perturbation theory, 
${\mathbf{R}}(\mu, E_\mathrm{L})$ is given by~\cite{hung2024qeraman}
\begin{equation}
    {\mathbf{R}}(\mu, E_\mathrm{L})=\sum_{\bm{k}}\sum_{i=f,n,n^\prime}
    \frac{\mathcal{D}_{fn^\prime}(\bm{k})\cdot \mathcal{M}_{n^\prime n}^\mu(\bm{k})\cdot \mathcal{D}_{ni}^{\dagger}(\bm{k})}
    {(E_\mathrm{L}-E_{ni}-\mathrm{i}\gamma_n)(E_\mathrm{L}-E_{n^\prime i}\pm\hbar\omega_\mu-\mathrm{i}\gamma_{n^\prime})},
    \label{eq:R_SRR}
\end{equation}
where the summation on electron wavevector, $\bm{k}$, is taken over the first BZ, and the summation on $i$ ($f$) is taken for the $i$-th initial (the $f$-th final) valence band at  $\bm{k}$. In order to recombine with the photo-excited electron with a hole at the $i$-th band, we set $i=f$. The labels of $n$ and $n^\prime$ denote the intermediate states of the $n$-th and $n^\prime$-th conduction bands. The energy terms $E_{ni} \equiv E_n(\bm{k}) - E_i(\bm{k})$ and  $E_{n^\prime i}\equiv E_{n^\prime}(\bm{k}) - E_i(\bm{k})$ are the transition energies from the initial state $|i\rangle$ to the intermediate state $|n\rangle$ ($|n^\prime\rangle$). $\gamma_{n}$ ($\gamma_{n^\prime})$ denotes the broadening factor of the resonance event, which is inversely proportional to the lifetime of the photo-excited electron at the intermediate state. $\mathcal{D}_{ni} \equiv \langle n | \nabla | i \rangle$ and $\mathcal{D}_{fn^\prime}\equiv \langle f | \nabla | n' \rangle$ are so-called dipole vector \cite{n950} or simply matrix elements of electron-photon interaction for the absorption and emission of photons in the Raman process, respectively. 
Tensor product of $\mathcal{D}_{ni}$ and $ \mathcal{D}_{fn^\prime}$ makes the expression of ${\mathbf{R}}$ the second rank tensor. 
$\mathcal{M}_{n^\prime n}^\mu$ is an electron-phonon matrix element, which describes the inelastic scattering of the excited electron in $n$ state to  $n^\prime$ by emitting the $\mu$-th phonon.

Resonance condition is obtained by either the first term ($E_\mathrm{L}-E_{ni}$) or the second term ($E_\mathrm{L}-E_{n^\prime i}\pm\hbar\omega_\mu$) in the energy denominator becomes zero, which corresponds to the incident or scattering resonance condition. The incident resonance condition is common to all the phonon modes, while the scattered resonance condition depends on the phonon energy. In the calculation, we further put the broadening factor to phonon frequency $\hbar\omega_\mu +i\Gamma_\mu$, in which $\Gamma_\mu$ is proportional to the inverse of the lifetime of the $\mu$-th phonon. $\Gamma_\mu$ gives the spectral width of Raman spectra.

\subsection{Double-resonance Raman intensity}
\label{sec:I-drr}
The DRR intensity as a function of $\omega_\text{RS}$ is similarly expressed by
\begin{eqnarray} 
\label{eq:DRR}
I\left(\omega_\text{RS}\right)&\propto&\sum_{\bm{q},\mu,\nu}\left|\mathbf{P}_s^{\dagger}\cdot{\mathbf{R}}\left(\bm{q},\mu,\nu, E_\mathrm{L}\right)\cdot\mathbf{P}_i\right|^2 \nonumber \\
&\,& \times\, \delta\left(\hbar\omega_\text{RS}\pm\hbar\omega_\mu\pm\hbar\omega_\nu\right) \\
&\equiv& \sum_{\bm{q},\mu,\nu} I^{\mu\nu}_{\bm{q}}, \nonumber
\end{eqnarray}
where the summation on $\bm{q}$ is taken over the first BZ.
In the case of two-phonon scattering, two independent phonon modes of the $\mu$-th and $\nu$-th phonon dispersions are considered with the phonon wavevectors of $-\bm{q}$ and $\bm{q}$, respectively. The photo-excited electron is first scattered from ($n$, $\bm{k}$) to ($n^\prime$, $\bm{k}+\bm{q}$) state by emitting the ($\mu$, $-\bm{q}$) phonon. Then, 
the electron is scattered from ($n^\prime$, $\bm{k}+\bm{q}$) to ($n^{\prime\prime}$, $\bm{k}$)
by emitting the ($\nu$, $\bm{q}$) phonon.
The Raman tensor for the DRR scattering takes a fourth-order perturbation form, which is given by
\newpage
\begin{strip}
\begin{equation}
    {\mathbf{R}}(\bm{q},\mu,\nu,E_\mathrm{L})=\sum_{\bm{k}}\sum_{i=f,n,n^\prime,n^{\prime\prime}}\frac{\mathcal{D}_{fn^{\prime\prime}}(\bm{k})\cdot \mathcal{M}_{n^{\prime\prime}n^\prime}^{\nu_{-\bm{q}}}(\bm{k}+\bm{q})\cdot \mathcal{M}_{n^{\prime\prime}n^\prime}^{\mu_{\bm{q}}}(\bm{k})\cdot \mathcal{D}_{ni}^{\dagger}(\bm{k})}{(E_\mathrm{L}-E_{ni}-\mathrm{i}\gamma_n)(E_\mathrm{L}-E_{n^\prime i}\pm\hbar\omega_\mu-\mathrm{i}\gamma_{n^\prime})(E_\mathrm{L}-E_{n^{\prime\prime}i}\pm\hbar\omega_\mu\pm\hbar\omega_\nu-\mathrm{i}\gamma_{n^{\prime\prime}})},
    \label{eq:R_DRR}
\end{equation}
\end{strip}
\noindent where $\pm$ can have four possible ways, including two-phonon Stokes (or anti-Stokes) Raman spectra and a pair of phonon-absorption and phonon-emission Raman spectra. 

For a two-phonon assignment in the DRR spectrum, the  contribution of Raman intensity from a pair of  $\mu$-th and $\nu$-th phonon modes, $I^{\mu\nu}$ to the total Raman intensity $I$ is given  by summing $I_{\bm{q}}^{\mu\nu}$ on $\bm q$ as follows:
\begin{equation}
\zeta_{\mu\nu} =\frac{\sum_{\bm{q}}I_{\bm{q}}^{\mu\nu}}{I}.
\label{eq:zeta}
\end{equation}
By taking the largest value of $\zeta_{\mu\nu}$, ($0 \le \zeta_{\mu\nu} \le 1$ ), we can obtain the assignment of DRR modes.

When we obtain Raman intensity $I_{\bm{q}}^{\mu\nu}$ as a function of $\bm{q}$ for a given pair of $\mu$ and $\nu$ from Eq.~\eqref{eq:DRR}, we can say which phonon wavevector $\bm{q}$ is the most relevant to the Raman peak from the value of $\zeta_{\mu\nu}$.   


\subsection{Defect-induced Raman intensity}
\label{sec:I-defect}
The defect-induced Raman spectrum is obtained by the DRR calculation.
In the case of one-phonon and one-elastic-scattering processes, one of two electron-phonon matrix elements in Eq.~\eqref{eq:R_DRR} is changed to the matrix element of the elastic scattering by the electron-defect coupling. The electronic scattering of a photo-excited electron from $\bm{k}$ to $\mathbf{k+q}$ occurs by $\bm{q}$ component of the Fourier transformed defect potential.  

When the defect potential is short compared with the lattice constant, a relatively large $\bm{q}$ component compared with the Brillouin zone is dominant. When the defect potential is long-range compared with the lattice constant, a relatively small $\bm{q}$ component is dominant. 
In the case of graphene, for example, the D band intensity becomes large when the defect potential is short range, such as an atomic defect or the armchair edge, while the D$^\prime$ band intensity becomes large when the defect potential is long range, such as ionic potential near the graphene plane.  

In the present version of {\sc QR$^2$-code}, we adopt the first approximation that the electron-defect matrix element of elastic scattering is a constant. Thus, we expect that both D and D$^\prime$ bands appear in the calculation. For a more explicit calculation of defect-induced Raman spectra, we need to calculate the Fourier transform of the defect potential for an electron in the material. The {\sc QR$^2$-code} is flexible for accepting the $\bm{q}$ dependence of elastic scattering matrix element (see below).

\section{Installation, workflow, and usage}
\label{sec:installandrun}
\begin{figure*}[t]
  \centering \includegraphics[width=17cm]{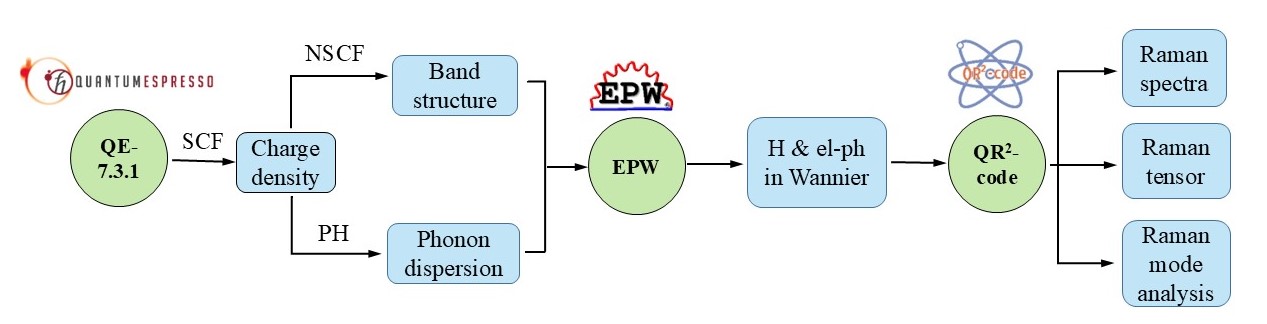}
  \caption{\label{fig:qeraman} Flowchart of the workflow to calculate the Raman spectra by green circles: (1) {\sc QE}, (2) {\sc EPW}, and (3) {\sc QR$^2$-code}, in which the figures from right to left are the logo of {\sc QE}, {\sc EPW}, and {\sc QR$^2$-code}, respectively. In blue rounded-rectangles, we show what we will calculate from the programs.}
\end{figure*}

In this section, we explain how to install and use {\sc QR$^2$-code}. The overall workflow of the calculations is shown in Fig.~\ref{fig:qeraman}.

\subsection{Download and installation}
\label{sec:install}
The current version of {\sc QR$^2$-code} is based on version 7.3.1 of {\sc QE}. Thus, first, we need to install {\sc QE} v7.3.1 for installing {\sc QR$^2$-code}. Note that the {\sc EPW} code, as a module of {\sc QE}, is automatically included in the {\sc QE} v7.3.1. Then, {\sc QR$^2$-code} is installed by the following steps:
\begin{enumerate}
   \item[(1)] Download {\sc QE} v7.3.1 from its official website: \href{https://www.quantum-espresso.org}{https://www.quantum-espresso.org}
    \item[(2)] Download the {\sc QR$^2$-code} at GitHub page:
    \href{https://github.com/JoeyyHuang/QR2-code}{https://github.com/JoeyyHuang/QR2-code}. 
    Alternatively, the readers can download the latest version of {\sc QR$^2$-code} directly from the GitHub repository by following the git command: 
    \begin{terminal2}{}
    *\$* git clone https://github.com/JoeyyHuang/QR2-code.git
    \end{terminal2}
    The code contains four directories \verb|PHonon-PH|, \verb|EPW-src|, \verb|phonon_sort| and \verb|Raman_PP|, and an \verb|examples| folder in the parent directory of \verb|QR2-code|.
    \item[(3)] Unpack {\sc QE} and {\sc QR$^2$-code} tarballs, and change to the directory of {\sc QE}. Then, copy the files in \verb|EPW-src| and \verb|PHonon-PH| of \textsc{QR$^2$-code} into the directory of \verb|EPW/src| and \verb|PHonon/PH| of {\sc QE}, respectively, and replace the original ones.
    \item[(4)] Change to the main directory of {\sc QE}, configure and compile by the general method of {\sc QE}. It is noted that the executable files of \verb|pw.x|, \verb|ph.x|, \verb|q2r.x|, \verb|matdyn.x| and \verb|epw.x| should be included. 
    You can compile them by following these commands:
    \begin{terminal2}{}
    *\$* ./configure
    *\$* make pw ph epw
    \end{terminal2}    
    If everything is done smoothly, these executables will be created in the \verb|bin| folder with the created date. It is noted that in the case that {\sc QE} already compiled before, the reader needs to recompile \verb|ph| and \verb|epw| again because we replace the original files in the directory of \verb|PHonon/PH| and \verb|EPW/src| of {\sc QE}. 
    \item[(5)] The files in \verb|phonon_sort| and \verb|Raman_PP| folders of the {\sc QR$^2$-code} are isolated codes. Copy them to the \verb|bin| folder of {\sc QE}, and compile by following commands:
    \begin{terminal2}{}
    *\$* ifort phonon_sort.f90 -o phonon_sort.x
    *\$* mpiifort raman_pp.f90 -o raman_pp.x
    \end{terminal2}
    Now, we have finished the installation of {\sc QR$^2$-code}.
    We list these executable files in Table~\ref{tab:command}.
\end{enumerate}
It is noted that we recommend using \verb|ifort| compiler. An alternative compiler is \verb|gfortran| and \verb|openMPI|, which are available on Linux.


\begin{table*}[ht!]
\centering
\caption{The commands in {\sc QR$^2$-code}.}
 \begin{tabular}{p{2.5cm} p{14cm}} 
 \toprule
 \textbf{Command} & \textbf{Purpose}\\ \midrule
 \texttt{epw.x} & Modified \texttt{epw.x} of {\sc QE}, the main executable to calculate the resonance Raman spectra.\\ \hdashline
  \texttt{phonon\_sort.x} & Categorize the phonon modes in a branch-resolved way over the whole BZ, which is useful for the assignment of DRR.\\ \hdashline
   \texttt{matdyn.x} & Modified \texttt{matdyn.x} of {\sc QE}, the core code to categorize the phonon mode for \texttt{phonon\_sort.x}.\\ \hdashline
 \texttt{raman\_pp.x} & Raman data post-processing code to generate the Raman spectrum in a Lorentzian shape and also optionally analyze the Raman bands in terms of phonon modes.\\
 \bottomrule
 \end{tabular}
\label{tab:command}
\end{table*}

\subsection{Workflow and usage}
\label{sec:run}
In Fig.~\ref{fig:qeraman}, we show the flowchart of the process to calculate the resonance Raman spectra $I(\omega_\text{RS})$. Six steps are required for the calculation of $I(\omega_\text{RS})$ with the commands as follows: 

\begin{itemize}
    \item[-] \textbf{Step 1}: First, the electron and phonon databases are obtained using {sc QE} by the following commands:
    \begin{terminal2}{}
    *\$* mpirun -np 56 pw.x < scf.in > scf.out
    *\$* mpirun -np 56 ph.x < ph.in > ph.out
    \end{terminal2}
    In the first line, we run \verb|pw.x| of {\sc QE} with 56 processors (the value of 56 can be changed manually) by the command \verb|mpirun| (running a program with parallel processors) with the input file \verb|scf.in| and the output file \verb|scf.out|. 
    In the second line, we run \verb|ph.x| of {\sc QE} with 
    56 processors with the input file \verb|ph.in| and the output file \verb|ph.out|. 
    In step 1, we obtain the database, including the band structure $E({\bm{k}})$, the wavefunctions $\psi_{\bm{k}}$, the lattice dynamical matrices (\verb|dyn*| files), and phonon deformation potential (\verb|dvscf*| files) in terms of a coarse $\bm{q}$ grid. 
    
    \item[-] \textbf{Step 2}: Next, we obtain phonon dispersions, which are needed in the double and defect-induced double resonance Raman calculations, as follows:
    \begin{terminal2}{}
    *\$* mpirun -np 56 q2r.x < q2r.in > q2r.out
    *\$* phonon_sort.x < phonon_sort.in
    \end{terminal2}
    In the first line, we run \verb|q2r.x| of {\sc QE} in 56 processors with the input file \verb|q2r.in| and the output file \verb|q2r.out|, which yields the interatomic force constants (\verb|ifc| file). In the second line, we run \verb|phnonon_sort.x| in a single processor with the input file \verb|phnonon_sort.in|. In step 2, the branch-ordered phonon modes are stored in a file named \verb|phonon_sort|.
    
    \item[-] \textbf{Step 3}: In this step, we will calculate electron-phonon matrix elements by using {\sc EPW} as follows:
    \begin{terminal2}{}
    *\$* python save.py
    *\$* mpirun -np 56 pw.x -npool 56 < nscf.in > nscf.out
    *\$* mpirun -np 56 epw.x -npool 56 < epw.in > epw.out
    \end{terminal2}
    In the first line, we run the python file \verb|save.py| to generate the folder \verb|save|, including the \verb|dyn*| and \verb|dvscf*| files from step 2, which are required for the \verb|epw.x| calculation.
    In the second line, run \verb|pw.x| with 56 processors for the non-SCF calculation with the input file \verb|nscf.in| and the output file \verb|nscf.out|. 
    It notes that the number of electron bands should be specified in \verb|nscf.in|, in which the number should be larger than one you want to Wannierize.
    In the third line, we run \verb|epw.x| with 56 processors for the Wannierization process with the input file \verb|epw.in| and the output file \verb|epw.out|.
    It is noted that the number of pool \verb|-npool| has to be the same as the total number of core \verb|-np|.    
    In step 3, the calculation implements the winterization process, i.e., obtaining the Hamiltonian, dynamical matrix, and electron-phonon matrix element in the Wannier basis.
    
    \item[-] \textbf{Step 4}: In this final step, we calculate the Raman spectra as follows:
    \begin{terminal2}{}
    *\$* mpirun -np 56 epw.x -npool 56 < epw-raman.in > epw-raman.out
    *\$* mpirun -np 56 raman_pp.x < raman_pp.in > raman_pp.out
    \end{terminal2}
    In the first line, we run \verb|epw.x| again with 56 processors for the resonance Raman calculation with the new input file \verb|epw-raman.in| and the output file \verb|epw-raman.out|. This calculation is actually a restarting \textsc{EPW} calculation on the breaking point of the last-step \textsc{EPW} performance. The input parameter for Raman calculation is added at the end of \verb|epw-raman.in|.
    In the second line, we run \verb|raman_pp.x| with 56 processors for the post-processing calculation with the input file \verb|raman_pp.in| and the output file \verb|raman_pp.out|.
    The post-processing code enables us to do two things. 
    The first one is to obtain the Raman spectra (having non-zero bandwidth) by a Lorentzian broadening on the basis of the Raman intensity data acquired in the last step calculation. 
    The second one is to analyze the (defect-induced) double resonance Raman band by providing the pair of phonon modes that give rise to the Raman band. 
    If the Raman spectra consist of a couple of phonon pairs, the percentage contribution of each pair to the overall Raman band intensity is provided in the output file \verb|raman_pp.out|. 
\end{itemize}

\section{Examples}
\label{sec:example}
In this section, we take two-dimensional (2D) graphene as an example to illustrate how to run the QR$^2$-code to obtain the resonance Raman spectra, including SRR, DRR, and defect resonance Raman. The input files used in the calculations and some standard output files can be found on GitHub (\href{https://github.com/JoeyyHuang/QR2-code/examples/Graphene}{QR$^2$-code/examples/Graphene}). We also provide other examples of MoS$_2$ monolayer and bulk h-BN on GitHub. Note that we have adopted relatively strict parameters in our calculations to ensure convergence and achieve better results. Therefore, reproducing the results of these calculations may take a considerable amount of time.

\subsection{Single-resonance Raman of graphene}
\label{sec:srr}
\noindent\ding{111}\enspace\textbf{Purpose:} The purpose is to calculate the linearly-polarized SSR spectra for several values of incident laser energy.

\noindent\ding{111}\enspace\textbf{How to run:} Follow steps 1, 3, and 4 to run this tutorial, as shown in Sec.~\ref{sec:run}. We do not need to run step 2 for SRR. It notes that the \verb|Graphene| directory includes five sub-folders: \verb|phonon|, \verb|epw|, \verb|raman|, \verb|phonon_sort|, and \verb|pseudo|. The \verb|pseudo| directory contains a pseudopotential file for the C atom. Step 1 will run in the \verb|phonon| directory, step 3 will run in the \verb|epw| directory, and step 4 will run on the \verb|raman/single| directory.

\noindent\ding{111}\enspace\textbf{Input files:} 
The input file \verb|scf.in| for step 1 is given as follows:
\begin{codeQEsmall}{examples/Graphene/phonon/scf.in}
&CONTROL
  calculation     = 'scf'
  prefix          = 'Gra'
  verbosity       = 'high'
  disk_io         = 'low'
  outdir          = '../tmp/'
  pseudo_dir      = '../pseudo/'
/
&SYSTEM
  ibrav           = 4
  celldm(1)       = 4.6608499868
  celldm(3)       = 8.1089334817
  nat             = 2
  ntyp            = 1
  ecutwfc         = 120
  ecutrho         = 480
  occupations     = 'smearing'
  smearing        = 'mv'
  degauss         = 0.01
  assume_isolated = '2D'
/
&ELECTRONS
  conv_thr        = 1.d-12
/
ATOMIC_SPECIES
  C  12.010  C.upf
ATOMIC_POSITIONS (crystal)
  C  0.3333333333  0.6666666667   0.0000000000
  C  0.6666666667  0.3333333333   0.0000000000
K_POINTS (automatic)
  30  30  1  0  0  0
\end{codeQEsmall}
The detailed explanation of the parameters in \verb|scf.in| is given in the hands-on guidebook of {\sc QE}~\cite{hung2022quantum} and on the web page: \url{https://www.quantum-espresso.org/Doc/INPUT\_PW.html}. It is noted that we use the Marzari-Vanderbilt smearing with a low broadening of degauss of 0.01 Ry for the occupation of a metallic system. Norm-conserving pseudopotential with the PBE exchange-correlation functional is used, which originates from the pseudo-Dojo distribution at \url{https://www.pseudo-dojo.org}. For a metallic system, a highly dense $\bm{k}$-point grid is usually necessary to properly catch the features at the Fermi surface. Here, the Monkhorst–Pack ${\bm k}$-point grid is chosen to be $30\times 30\times 1$ after a convergence test by our side. If the calculation is successfully completed, a \texttt{tmp} folder is generated containing the files of the ground-state charge density and wavefunctions for graphene. These files will be used in the next step for the phonon calculation.

The input file \verb|ph.in| for step 1 is given as follows:
\begin{codeQEsmall}{examples/Graphene/phonon/ph.in}
phonon
&inputph
  prefix    = 'Gra'
  outdir    = '../tmp/'
  nmix_ph   = 20
  tr2_ph    = 1.0D-15
  verbosity = 'high'
  reduce_io = .true.
  fildyn    = 'dyn'
  fildvscf  = 'dvscf'
  ldisp     = .true.
  nq1       = 15 
  nq2       = 15 
  nq3       = 1
/
\end{codeQEsmall}
The detailed explanation of the parameters in \verb|ph.in| is given in the hands-on guidebook of {\sc QE}~\cite{hung2022quantum} and on the web page: \url{https://www.quantum-espresso.org/Doc/INPUT_PH.html}.
Please note that the phonon calculation typically requires a very large amount of disk space for wavefunction-related data storage, especially when employing norm-conserving pseudopotentials. Here, we activate the \verb|reduce_io = .true.| at line 8 to substantially reduce the storage usage by keeping only the minimum data essential for subsequent calculations. At line 5,  \verb|nmix_ph = 20| is set to speed up convergence at the cost of using more memory. A $15\times 15\times 1$ $\bm{q}$-grid will generate 27 inequivalent q points within the BZ of graphene. After a successful calculation, there will be 27 dynamical matrix files named \verb|dyn1|, \verb|dyn2|, up to \verb|dyn27|, and lattice deformation potential files \verb|dvscf_q*| for each inequivalent q point can be found in the \verb|/tmp/_ph0/| directory.

The input file \verb|nscf.in| for step 2 is given as follows:
\begin{codeQEsmall}{examples/Graphene/epw/nscf.in}
&CONTROL
  calculation     = 'nscf'
  prefix          = 'Gra'
  verbosity       = 'high'
  disk_io         = 'low'
  outdir          = '../tmp/'
  pseudo_dir      = '../pseudo/'
/
&SYSTEM
  ibrav           = 4
  celldm(1)       = 4.6608499868
  celldm(3)       = 8.1089334817
  nat             = 2
  ntyp            = 1
  nbnd            = 15
  ecutwfc         = 120
  ecutrho         = 480
  occupations     = 'smearing'
  smearing        = 'mv'
  degauss         = 0.01
  assume_isolated = '2D'
/
&ELECTRONS
  conv_thr        = 1.d-12
  diago_full_acc  = .true.
  diago_thr_init  = 1.d-9
/
ATOMIC_SPECIES
  C  12.010  C.upf
ATOMIC_POSITIONS (crystal)
  C  0.3333333333  0.6666666667   0.0000000000
  C  0.6666666667  0.3333333333   0.0000000000
K_POINTS (crystal)
  900
  0.0000000  0.0000000  0.0000000  1.111111e-03
  0.0000000  0.0333333  0.0000000  1.111111e-03
  0.0000000  0.0666667  0.0000000  1.111111e-03
...
\end{codeQEsmall}
Compared with the \verb|scf| calculation, the number of electronic bands, \verb|nbnd = 15|, is given at line 15. The list of $\bm k$-points at the namelist \verb|K_POINTS| corresponds to a uniform-grid of $30\times 30 \times 1$ with 900 lines, which can be created by the script file \verb|kmesh.pl| in the \verb|epw| directory by command as follows:
\begin{terminal2}{}
    *\$* ./kmesh.pl 30 30 1 > k-list.dat
\end{terminal2}
Then, the reader can copy the list of $\bm k$-points in \verb|k-list.dat| and paste it into \verb|nscf.in|.

The input file \verb|epw.in| for step 2 is given as follows:
\begin{codeQEsmall}{examples/Graphene/epw/epw.in}
epw
&inputepw
  prefix       = 'Gra'
  outdir       = '../tmp/'
  dvscf_dir    = './save/'
  system_2d    = 'dipole_sp'
  asr_typ      = 'simple'
  vme          = 'dipole'

  nbndsub      = 5
  ep_coupling  = .true.
  elph         = .true.
  use_ws       = .true.
  num_iter     = 400

  dis_froz_min = -25
  dis_froz_max = -1

  proj(1) = 'f=0.33333333,0.66666667,0.0:sp2;pz'
  proj(2) = 'f=0.66666667,0.33333333,0.0:pz'

  fsthick      = 5
  eps_acustic  = 5
  degaussw     = 0.005
  degaussq     = 0.01

  wdata(1)     = 'dis_mix_ratio = 0.7'
  wdata(2)     = 'dis_num_iter = 400'
  wdata(3)     = 'conv_window = 3'
  wdata(4)     = 'trial_step = 1.0'
  wdata(5)     = 'guiding_centres=.true.'

  epwwrite     = .true.
  epwread      = .false.
  wannierize   = .true.

  nk1  = 30, nk2  = 30, nk3  = 1
  nq1  = 10, nq2  = 10, nq3  = 1
  
  nkf1 = 1,  nkf2 = 1,  nkf3 = 1
  nqf1 = 1,  nqf2 = 1,  nqf3 = 1
 /
\end{codeQEsmall}
The detailed explanation of the parameters in \verb|epw.in| is given on the web page: \url{https://docs.epw-code.org/doc/Inputs.html}. It notes that the readers should be sure to run the python file \verb|save.py| before running \verb|epw.x| since the folder \verb|save| at line 5 is generated by \verb|save.py|.
If the calculation normally finishes, 
\verb|crystal.fmt|, \verb|dmedata.fmt|, \verb|epwdata.fmt|, and \verb|Gra.ukk| files are obtained in the \verb|epw| directory, which will be used to run the subsequent Raman calculation in step 4.

The input file \verb|epw-raman.in| in step 4 is given as follows:
\begin{codeQEsmall}{examples/Graphene/raman/single/epw-raman.in}
epw
&inputepw
  prefix       = 'Gra'
  outdir       = '../../tmp/'
  dvscf_dir    = '../../epw/save/'
  system_2d    = 'dipole_sp'
  asr_typ      = 'simple'
  vme          = 'dipole'

  nbndsub      = 5
  ep_coupling  = .true.
  elph         = .true.
  use_ws       = .true.
  num_iter     = 400

  fsthick      = 5
  eps_acustic  = 5
  degaussw     = 0.005
  degaussq     = 0.01

  epwwrite     = .false.
  epwread      = .true.
  wannierize   = .false.

  nk1  = 30,   nk2  = 30,   nk3  = 1
  nq1  = 15,   nq2  = 15,   nq3  = 1
  
  nkf1 = 150,  nkf2 = 150,  nkf3 = 1
  nqf1 = 1,    nqf2 = 1,    nqf3 = 1

  Raman_type    = 'single'
  lhoph         = .true.
  Elaser        = 1.96
  Egamma        = 0.1
  reson_lim     = .true.
  reson_thr     = 1
  prtdipole     = .true.
  prtRaman      = .true.
  polar         = 'all'
  filraman      = 'qraman'
  ltensor       = .true.
 /
\end{codeQEsmall}
The meaning of all parameters related to the Raman calculation (from lines 31 to 41) is listed in Table~\ref{tab:raman}.
The settings of \verb|epwwrite = .false.| and \verb|epwread = .true.| at lines 20 and 21, respectively, indicate that we are actually performing a restarting calculation of \verb|epw.x| based on the Wannier basis that we get in step 2. The setting \verb|wannierize = .false.| at line 22 indicates that the Wannier function is read from \verb|Gra.ukk| file. Thus, please ensure that the four files, including \verb|crystal.fmt|, \verb|dmedata.fmt|, \verb|epwdata.fmt|, and \verb|Gra.ukk|, are copy and paste from the \verb|epw| directory to the \verb|single| directory. 
At line 27, we set \verb|nkf1 = 150, nkf2 = 150, nkf3 = 1| for interpolation a dense $\bm{k}$-grid of $150\times 150 \times1$ in the Raman calculation. On the other hand, we set \verb|nqf1 = 1, nqf2 = 1, nqf3 = 1| at line 29 for the SRR calculation to consider only the phonon at $\Gamma$ point. 
The SRR calculation is set by \verb|Raman_type = 'single'| at line 31. The laser energy $E_\mathrm{L}$ and the broadening factor $\gamma$ in Eq.~\eqref{eq:R_SRR} are set by \verb|Elaser = 1.96| and \verb| Egamma = 0.1| in the unit of eV at lines 33 and 34, respectively. If the calculation normally finishes, the Raman intensity is written in the output files \verb|qraman*|.

\begin{table*}[t]
\centering
\caption{Meaning of input variables in \texttt{epw-raman.in} file for Raman calculation.}
 \begin{tabular}{p{0.4cm} p{2.4cm} p{14cm}} 
 \toprule
 \textbf{Line} &\textbf{Syntax} & \textbf{Meaning}\\ \midrule
 1& \texttt{prtRaman} & \texttt{.true.} is set, if printing Raman intensity. The default value is \texttt{.true.} \\ \hdashline
 2& \texttt{Raman\_type} & Type of calculated resonance Raman. \texttt{`single'}: single-resonance Raman, \texttt{`double'}: double-resonance Raman, \texttt{`defect'}: defect-induced double-resonance Raman. The default value is \texttt{`double'}.\\ \hdashline
 3& \texttt{lhoph} & \texttt{.true.} is set, if considering the hole-phonon coupling. The default value is \texttt{.false.}\\ \hdashline
 4& \texttt{reson\_lim} & \texttt{.true.} is set, if taking a limitation on the energy level separation in resonance transition. The default value is \texttt{.true.}\\ \hdashline
 5& \texttt{reson\_thr} & Set the threshold value of limitation, if  \texttt{reson\_lim = .true.}. The default value is \texttt{1.d0.}\\ \hdashline
 6& \texttt{Elaser} & The incident laser excitation energy in the unit of eV. The default value is \texttt{1.96d0}.\\ \hdashline
 7& \texttt{Egamma} &  The electronic energy broadening in the unit of eV, relevant to the electron life. The default value is \texttt{1.d-1}.\\ \hdashline
 8& \texttt{polar} & Polarization of Raman scattering. \texttt{`custom'}: calculate the Raman in given polarization, \texttt{`all'}: output all possible polarization. The default value is \texttt{`custom'.}\\ \hdashline
 9& \texttt{filraman} & Set the file name where the Raman intensity is stored, if \texttt{polar = `custom'}. The default value is \texttt{`qraman'}\\ \hdashline
 10& \texttt{ei} & Set incident light polarization, if \texttt{polar = `custom'}. The default value is \texttt{(/ (1.d0,0.d0), (0.d0,0.d0), (0.d0,0.d0) /)}.\\ \hdashline
 11& \texttt{es} & Set scattered light polarization, if \texttt{polar = `custom'}. The default value is \texttt{(/ (1.d0,0.d0), (0.d0,0.d0), (0.d0,0.d0) /)}.\\ \hdashline
 12& \texttt{Cq} & The coefficients of each power of q in formula of $\bm{q}$-dependent defect-electron scattering. The default value is \texttt{ (/ 0.d0, 0.d0, 0.d0, 1.d0, 0.d0, 0.d0, 0.d0 /)}.\\ \hdashline
 13& \texttt{lBE} & \texttt{.true.} is set, if considering the temperature in the unit of K at the phonon Bose–Einstein distribution. The default value is \texttt{.false.}\\ \hdashline
 14& \texttt{temphon} & Set the temperature in the unit of K at phonon Bose–Einstein distribution, if \verb|lBE=.true.|. The default value is \texttt{3.d2.}\\ \hdashline
 15& \texttt{ltensor} & \texttt{.true.} is set, if printing Raman tensor. The default value is \texttt{.false.}\\ \hdashline
 16& \texttt{qtensor\_start} & Set the first $\bm{q}$ of which the Raman tensor is printed, if \verb|ltensor = .true.|. The default value is \texttt{1.}\\ \hdashline
 17& \texttt{qtensor\_end} & Set the last $\bm{q}$ of which the Raman tensor is printed, if \verb|ltensor = .true.|. The default value is \texttt{1.}\\ \hdashline
 18& \texttt{prtdipole} & \texttt{.true.} is set, if printing the interpolated dipole matrix elements. The default value is \texttt{.true.}\\
 \bottomrule
 \end{tabular}
\label{tab:raman}
\end{table*}

The input file \verb|raman_pp.in| in step 4 is given as follows:
\begin{codeQEsmall}{examples/Graphene/raman/single/raman\_pp.in}
&PLOT
  dir_raman      = './qraman-xx'
  Raman_type     = 'single'
  lhoph          = .true.
  lRay_sca       = .false.
  Ray_thr        = 5
  Rs_min         = 0
  Rs_max         = 2000
  Rs_inc         = 0.1
  Lgamma         = 20
/
\end{codeQEsmall}
The parameters from lines 2 to 10 are related to the post-processing Raman spectra and are explained in detail in Table~\ref{tab:raman-pp}. The parameter \verb|Lgamma = 20| gives the broadening parameter in cm$^{-1}$. If the calculation normally finishes, the Raman spectra are written in the output file \verb|Raman.dat|. 

\begin{table*}[t]
\centering
\caption{Meaning of input variables in \texttt{raman\_pp.in} file.}
 \begin{tabular}{p{0.4cm} p{2.4cm} p{14cm}} 
 \toprule
 \textbf{Line} &\textbf{Syntax} & \textbf{Meaning}\\ \midrule
 1& \texttt{dir\_raman} & The file where the Raman spectra are stored in the previous run. The default value is \texttt{../qraman}. \\ \hdashline
 2& \texttt{Raman\_type} & Type of calculated resonance Raman. \texttt{`single'}: single-resonance Raman, \texttt{`double'}: double-resonance Raman, \texttt{`defect'}: defect-induced double-resonance Raman. It must keep the same set as in \texttt{epw-raman.in}. The default value is \texttt{`double'}.\\ \hdashline
 3& \texttt{lhoph} & It must keep the same set as in \texttt{epw-raman.in}. The default value is \texttt{.false.}\\ \hdashline
 4& \texttt{lRay\_sca} & If \texttt{.true.}, consider the Rayleigh scattering. It is recommended to set \texttt{.false.} in order to undermine the strong Raman intensity in the vicinity of 0 cm$^{-1}$. The default value is \texttt{.false.}\\ \hdashline
 5& \texttt{Ray\_thr} & If \texttt{lRay\_sca = .false.}, this set the radius threshold centered at 0 cm$^{-1}$, within which the Raman intensity is undermined. The default value is \texttt{5.d0}\\ \hdashline
 6& \texttt{Rs\_min} & The minimum frequency in Raman shift in the unit of cm$^{-1}$. The default value is \texttt{-4.d3}.\\ \hdashline
 7& \texttt{Rs\_max} &  The maximum frequency in Raman shift in the unit of cm$^{-1}$. The default value is \texttt{4.d3}.\\ \hdashline
 8& \texttt{Rs\_inc} & the frequency increment in Raman shift in the unit of cm$^{-1}$. The default value is \texttt{1.d-1}.\\ \hdashline
 9& \texttt{Lgamma} & Lorentz function broadening, relevant to the FWHM of Raman peaks. The default value is \texttt{1.d1}.\\ \hdashline
 10& \texttt{lRaman\_modes} & If \texttt{.true.}, perform Raman modes analysis. This, along with the following parameters, only works in (defect-induced) double resonance Raman cases. The default value is \texttt{.false.}\\ \hdashline
 11& \texttt{nphonon\_modes} & How many pairs of phonon modes are demonstrated for the assignment of each Raman band. The default value is \texttt{6}.\\ \hdashline
 12& \texttt{nRaman\_modes} & How many Raman modes are analyzed. The default value is \texttt{1}.\\ \hdashline
 13& \texttt{Raman\_modes(:)} & The Raman shifts of \texttt{nRaman\_modes} Raman modes that are to be analyzed. The default value is \texttt{0.d0}.\\
 \bottomrule
 \end{tabular}
\label{tab:raman-pp}
\end{table*}

\noindent\ding{111}\enspace\textbf{Output files:}

\begin{figure}[t]
  \centering 
  \includegraphics[width=0.9\linewidth]{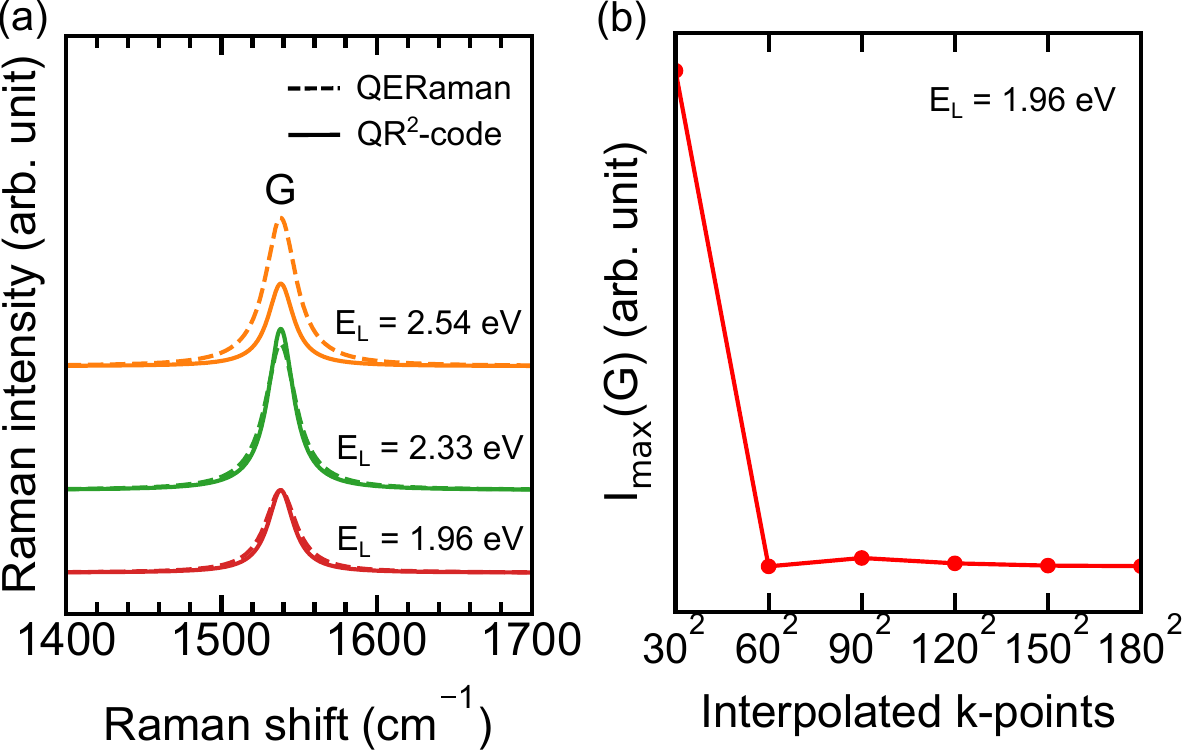}
  \caption{\label{fig:SRR} (a) Single-resonance Raman spectra of graphene at laser energies of 1.96, 2.33, and 2.54 eV, in which solid and dashed lines are calculated by {\sc QR$^2$-code} and {\sc QERaman}, respectively. 
  (b) The convergence test of the maximum Raman intensity as a function of interpolated ${\bm k}$-points for 1.96 eV laser energy.}
\end{figure}

The Raman spectra can be found in the output file \verb|Raman.dat| in Sec.~\ref{sec:srr} which are calculated for $E_\mathrm{L}=1.96$ eV. To calculate the Raman spectra for other laser energies, the reader needs to modify the value of \verb|Elaser| at line 33 of \verb|epw-raman.in| and then perform the calculation from step 4. In Fig.~\ref{fig:SRR}(a), we plot the Raman spectra of graphene for $E_\mathrm{L}=1.96$, 2.33, and 2.54 eV, which show the G peak at 1540 cm$^{-1}$, which reproduces the observed G peak ($\sim 1580$ cm$^{-1}$) in the experiment~\cite{ferrari2006raman}. It is noted that the format of \verb|Raman.dat| includes four columns: 1st column is the Raman shift $\omega_\text{RS}$, 2nd column is total Raman spectra, and 3rd and 4nd columns are Raman spectra with different Raman scattering process pathways, including the $e$-process (the intermediate state at conduction band) and $h$-process (the intermediate state at valence band). These processes are essential for the DRR. However, for the SRR, we consider only the case of the $e$-process in Fig.~\ref{fig:SRR}, which means that the excited electron is scattered in the conduction band.

Compared with {\sc QERaman}~\cite{hung2024qeraman} (dashed line), {\sc QR$^2$-code} shows similar SRR spectra at the $E_\mathrm{L}=1.96$ and 2.33 eV. However, at higher laser energy $E_\mathrm{L}=2.54$ eV, {\sc QERaman} shows larger Raman intensity than that of {\sc QR$^2$-code} as shown in Fig.~\ref{fig:SRR}(a). This might come from the contribution of the many bands to the resonance conditions in {\sc QERaman}. It is noted that the energy band structure of {\sc QERaman} is directly based on the density functional theory (DFT) calculations in {\sc QE}, while {\sc QR$^2$-code} is based on the Wannier interpolation in {\sc EPW}, which considers fewer bands than DFT. Nevertheless, the Wannier interpolation is an important method to speed up DRR calculations, as discussed in Sec.~\ref{sec:drr}.

In Fig.~\ref{fig:SRR}(b), we plot the maximum intensity of the G band, $I_{\max}(\text{G})$, for $E_\mathrm{L} = 1.96$ eV as a function of the number of $\bm k$-grid (see line 28 in \verb|epw-raman.in|). The convergence value of $\bm k$-grid related to $I_{\max}(\text{G})$ can be found at $60\times 60\times1$ or higher.

\subsection{Double-resonance Raman of graphene}
\label{sec:drr}
\noindent\ding{111}\enspace\textbf{Purpose:} The purpose is to calculate the linearly-polarized DRR spectra for a given value of incident laser energy.

\noindent\ding{111}\enspace\textbf{How to run:} Follow the four steps to run this paper, as shown in Sec.~\ref{sec:run}, in which steps 1 and 3 are already done in Sec.~\ref{sec:srr}. The reader needs to run step 2 on the \verb|phonon_sort| directory and step 4 on the \verb|raman/double| directory. The reader also needs to copy and paste the files \verb|crystal.fmt|, \verb|dmedata.fmt|, \verb|epwdata.fmt|, and \verb|Gra.ukk| (in step 3) from the \verb|epw| directory to the \verb|raman/double| directory.

\noindent\ding{111}\enspace\textbf{Input files:} 
For step 2, the input file \verb|q2r.in| is given as follows:
\begin{codeQEsmall}{examples/graphene/phonon\_sort/q2r.in}
&input
  fildyn   = '../phonon/dyn'
  flfrc    = 'ifc'
  zasr     = 'simple'
  loto_2d  = .true.
/
\end{codeQEsmall}
The detailed explanation of the parameters in \verb|q2r.in| is given in the hands-on guidebook of {\sc QE}~\cite{hung2022quantum} or on the web page: \url{https://www.quantum-espresso.org/Doc/INPUT_Q2R.html}. At line 5, we used \verb|loto_2d = .true.| to activate two-dimensional treatment of LO-TO splitting. The output file containing the interatomic force constant in real space is named as \verb|ifc|, which is needed for \verb|phonon_sort.x|.

The input file \verb|phonon_sort.in| in step 2 is given as follows:
\begin{codeQEsmall}{examples/graphene/phonon/phonon\_sort.in}
&q_grid
  nqf1        = 150  
  nqf2        = 150 
  nqf3        = 1
  dql         = 0.005
  density     = 500
  flfrc       = './ifc'
  asr         = 'simple'
  loto_2d     = .true.
  matdyn_path = 'matdyn.x'
/
\end{codeQEsmall}
Lines 1-3 specifies the interpolation dense $\bm q$-grid of $150 \times 150 \times 1$, which will be used in the \verb|epw-raman.in| in the next step. \verb|dpl| at line 5 is the starting distance (in crystalline
coordinates) away from the $\Gamma$ point, and \verb|density| at line 6 is the number density of $\bm q$-points for obtaining the order of phonon dispersion correctly. \verb|phonon_sort.x| reads the \verb|ifc| file, automatically generates \verb|matdyn.in| files, and runs a loop calculation with \verb|matdyn.x| to obtain the correct order of phonon dispersion, which is stored in the output file \verb|phonon_sort|. 
If the matdyun.x does not work, the reader can check if (1) the file \verb|ifc| exists in the \verb|phonon| directory, and (2) the modified \verb|matdyn.x| exists in the folder specified by the PATH environment variable. Lines 8 to 9 are the options for \verb|matdyn.x|. If the calculation normally finishes, the output file \verb|phonon_sort| is created in the \verb|phonon| directory. The reader needs to copy the file \verb|phonon_sort| from the \verb|phonon| directory to the \verb|raman/double| directory manually.

\begin{figure*}[t]
  \centering 
  \includegraphics[width=0.75\linewidth]{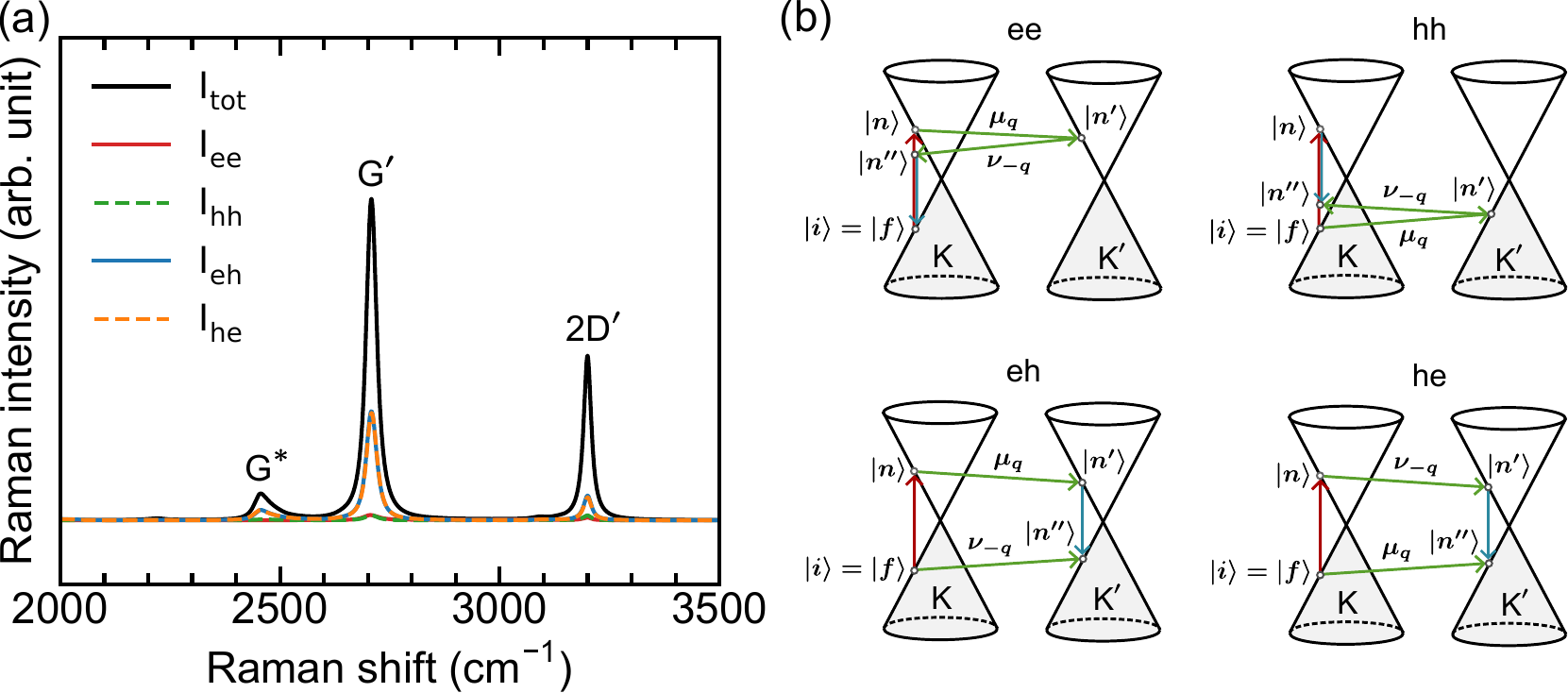}
  \caption{\label{fig:DRR1} (a) Double-resonance Raman spectra (DRR) of graphene at 1.96 eV laser energy, in which three DRR peaks correspond to G$^*$,  G$^\prime$, and 2D$^\prime$ modes.
  (b) Schematics of the DRR scattering pathways, including $ee$, $hh$, $eh$, and $he$, with either electron or hole or both involved in the carrier-phonon scattering process.}
\end{figure*}

For step 4, the input file \verb|epw-raman.in| is given as follows:
\begin{codeQEsmall}{examples/graphene/raman/double/epw-raman.in}
epw
&inputepw
  prefix       = 'Gra'
  outdir       = '../../tmp/'
  dvscf_dir    = '../../epw/save/'
  system_2d    = 'dipole_sp'
  asr_typ      = 'simple'
  vme          = 'dipole'

  nbndsub      = 5
  ep_coupling  = .true.
  elph         = .true.
  use_ws       = .true.
  num_iter     = 400

  fsthick      = 5
  eps_acustic  = 5
  degaussw     = 0.005
  degaussq     = 0.01

  epwwrite     = .false.
  epwread      = .true.
  wannierize   = .false.

  nk1  = 30,   nk2  = 30,   nk3  = 1
  nq1  = 15,   nq2  = 15,   nq3  = 1
  
  nkf1 = 150,  nkf2 = 150,  nkf3 = 1
  nqf1 = 150,  nqf2 = 150,  nqf3 = 1

  Raman_type    = 'double'
  lhoph         = .true.
  Elaser        = 1.96
  Egamma        = 0.1
  reson_lim     = .true.
  reson_thr     = 1
  prtdipole     = .true.
  prtRaman      = .true.
  polar         = 'all'
  filraman      = 'qraman'
 /
\end{codeQEsmall}
Since the DRR involves the phonon modes throughout the Brillouin zone, the fine $\bm q$-points is specified in a dense grid of $150\times 150 \times 1$ at line 29. A dense grid of $150\times 150 \times 1$ is also set at line 28 for $\bm k$-points based on the convergence test in Sec. 4.1. \verb|Raman_type = 'double'| is set at line 31 to perform DRR calculation. We also consider the hole-phonon scattering in the Raman scattering process by setting \verb|lhoph = .true.| at line 32. 
After the calculation, the Raman intensity data is written in the file \verb|qraman*|.

It is important to note that, in step 4, we need to run \verb|raman_pp.x| twice, in which the first run uses \verb|raman_pp1.in| without \verb|&ANALYSIS|, and then the second run uses \verb|raman_pp2.in| with \verb|&ANALYSIS|. The input files \verb|raman_pp1.in| and \verb|raman_pp2.in| in step 4 is given as follows:

\begin{codeQEsmall}{examples/Graphene/raman/double/raman\_pp1.in}
&PLOT
  dir_raman      = './qraman-xx'
  Raman_type     = 'double'
  lhoph          = .true.
  lRay_sca       = .false.
  Ray_thr        = 5
  Rs_min         = 0
  Rs_max         = 4000
  Rs_inc         = 0.1
  Lgamma         = 20
/
\end{codeQEsmall}
Lines 2 to 10 are related to the post-processing Raman spectra and explained in detail in Table~\ref{tab:raman-pp}. Here, we select the $\overline{\text{Z}}$(XX)Z configuration by setting \verb|dir_raman = './qraman-xx'| at line 2. For the helicity-dependent Raman spectra, the readers can select \verb|qraman-ll| and \verb|qraman-lr| for $\sigma+\sigma+$ and $\sigma+\sigma-$, respectively, in which $\sigma+$ and $\sigma-$ denote left- and right-handed circularly-polarized light.

\begin{codeQEsmall}{examples/Graphene/raman/double/raman\_pp2.in}
&PLOT
  dir_raman      = './qraman-xx'
  Raman_type     = 'double'
  lhoph          = .true.
  lRay_sca       = .false.
  Ray_thr        = 5
  Rs_min         = 0
  Rs_max         = 4000
  Rs_inc         = 0.1
  Lgamma         = 20
/
&ANALYSIS
  lRaman_modes   = .true.
  nphonon_modes  = 6
  nRaman_modes   = 3
  Raman_modes(1) = 2456.1
  Raman_modes(2) = 2707.8
  Raman_modes(3) = 3199.7
/
\end{codeQEsmall}
Lines 13 to 18 are explained in detail in Table~\ref{tab:raman-pp}. We set \verb|nphonon_modes = 6| at line 14 for the six phonon modes for graphene. At line 15, we set \verb|nRaman_modes = 3| to calculate the two-phonon assignment in Eq.~\eqref{eq:zeta} for three DRR peaks, which are listed at lines 16-18. 

\noindent\ding{111}\enspace\textbf{Output files:}
The Raman spectra in the output file \verb|Raman.dat| are calculated for the laser energy of 1.96 eV, in which \verb|Raman.dat| includes six columns: 1st column is the Raman shift $\omega_\text{RS}$, 2nd column is the total Raman intensity $I_\text{tot}$, and 3rd, 4th, 5th and 6th columns are the contribution of Raman intensities ($I_{ee}$, $I_{hh}$, $I_{eh}$, and $I_{he}$) with different Raman scattering processes, including the $ee$-, $hh$-, $eh$-, and $he$-processes, respectively. In Figs.~\ref{fig:DRR1}(a) and (b), we plot the DRR intensity as a function of the Raman shift and the schematics of the Raman scattering processes, respectively. Three DRR peaks are found at 2456.1, 2707.8, and 3199.7 cm$^{-1}$, which are consistent with the observed DRR peaks, G$^*$ (or D+D$^{\prime\prime}$), G$^\prime$ (or 2D), and 2D$^\prime$, in the experiment~\cite{liu2015deep,cong2018stokes}.

As shown in Fig.~\ref{fig:DRR1}(a), the DRR spectra are mainly dominated by the $eh$- and $he$-processes, in which the photon absorption and emission happen at different wave vector $\bm{k}$ (see Fig.~\ref{fig:DRR1}(b) bottom). Both electrons and the holes are involved in the DRR scattering in $eh$- and $he$-processes, while only electrons (or holes) are involved in the $ee$-process (or $hh$-process) that is killed by the destructive quantum interference~\cite{venezuela2011theory}. Thus, the dominating DRR scattering pathways are the $eh$- and $he$-processes at the laser energy $E_\mathrm{L} =1.96$ eV. On the other hand, Zhang \textit{et al.}~\cite{zhang2025duv} recently reported that for $4 < E_\mathrm{L} < 5$ eV (or the deep UV region),  the $ee$-process starts to dominate over other processes. This suggests that the interference effect depends on $E_\mathrm{L}$, which can be tuned at line 34 in \verb|epw-raman.in|.

\begin{figure}[t]
  \centering 
  \includegraphics[width=0.8\linewidth]{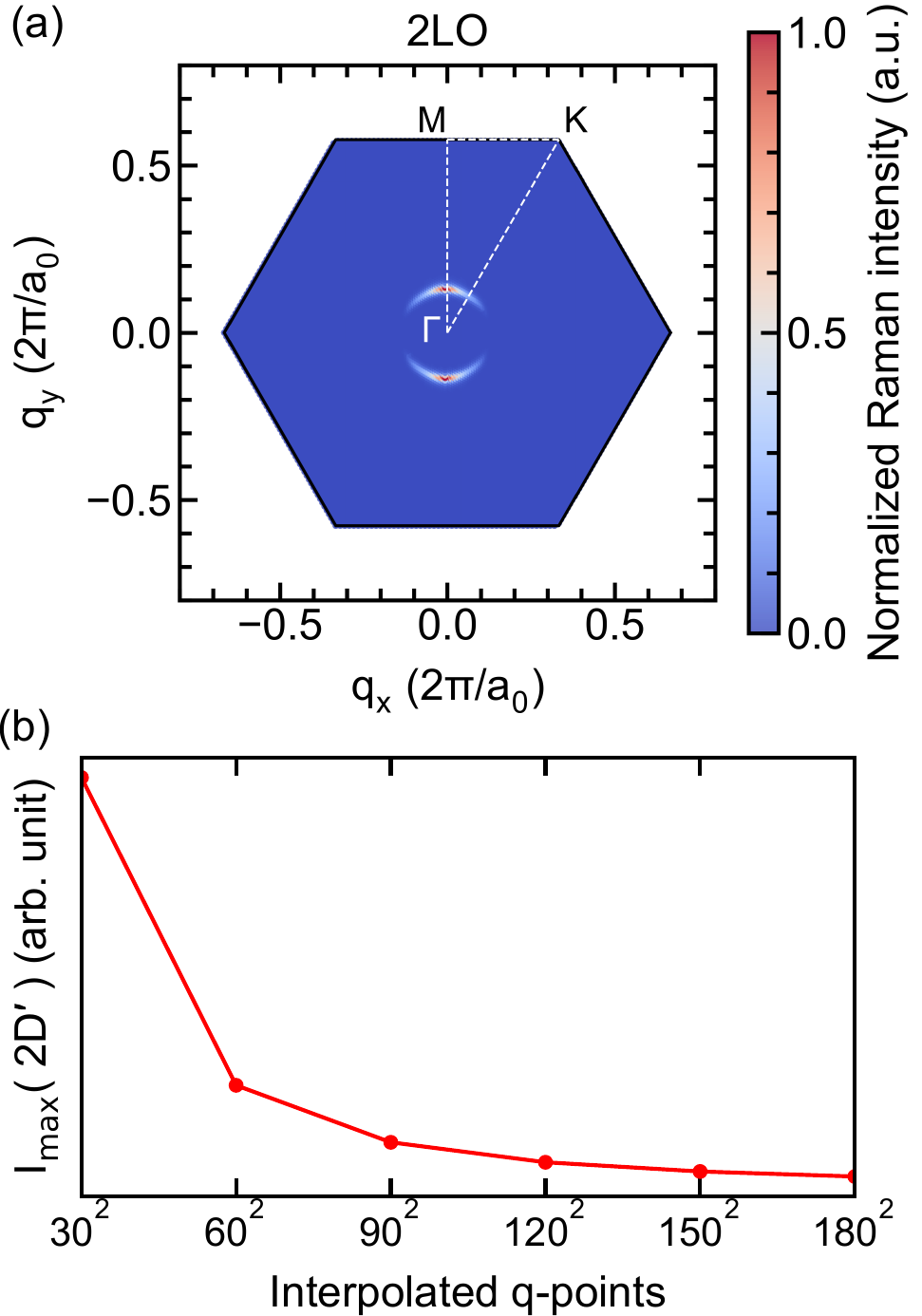}
  \caption{\label{fig:DRR2} (a) $\bm q$-dependent Raman intensity for the 2LO combinational mode for the 2D$^\prime$ band at 1.96 eV laser energy. Here, $q_x$ and $q_y$ are in units of $2\pi/a_0$, where $a_0$ is the lattice constant of graphene. 
  (b) The convergence test of the maximum Raman intensity of the 2D$^\prime$ band as a function of interpolated ${\bm q}$-points.}
\end{figure}

The assignment of the phonon modes for the DRR peaks can be found in the output file \verb|raman_pp2.out|. For the G$^*$ band, the main contributions of two-phonon modes are the combination of TO+LA (62.6\%) and LA+TO (33.7\%). It is noted that for graphene, the phonon orders from 1st to 6th correspond to ZA, TA, LA, ZO, TO, and LO, respectively~\cite{jorio2011raman}. The difference in contribution between TO+LA and LA+TO is due to the time-order phonon scattering processes~\cite{zhang2025duv}, in which the scattered ($n^\prime$, $\bm{k}+\bm{q}$) state (see Fig.~\ref{fig:DRR1} (b)) is located at a difference energy. Other combination of 2TO (62.4\%) and 2LO (35.3\%) are found for the G$^\prime$ band and 2LO (99.9\%) for the 2D$^\prime$ band. As an example, in Fig.~\ref{fig:DRR2}(a), we show the wavenumber $\bm{q}$-dependence of Raman intensity of the 2LO combinational mode for the 2D$^\prime$ band at $E_\mathrm{L}=1.96$ eV. The dominating contributed $\bm{q}_{\max}$ points around $\bm{q}_{\max}=(0, 0.1)$ (in units of $2\pi/a_0$,  where $a_0$ is the lattice constant of graphene). For the DRR, $\bm{q}_{\max}$ should depend on the $E_\mathrm{L}$~\cite{zhang2025duv}.

In Fig.~\ref{fig:DRR2}(b), we show the maximum value of Raman intensity of the 2D$^\prime$ band, $I_{\max}(\text{2D}^\prime)$, for $E_\mathrm{L} = 1.96$ eV as a function of the interpolated $\bm q$-grid (see line 29 in \verb|epw-raman.in|). The convergence value of interpolated $\bm q$-grid can be found at $150\times 150\times1$. It is noted that the interpolated $\bm k$-grid is set $150 \times 150 \times 1$ for all the $\bm q$-grid cases.

\subsection{defect-induced Raman of graphene}
\label{sec:defect}
\noindent\ding{111}\enspace\textbf{Purpose:} The purpose is to calculate the defect-induced Raman spectra as a function of incident laser energy.

\noindent\ding{111}\enspace\textbf{How to run:} Since steps 1, 2, and 3 are already done in Sec.~\ref{sec:drr}, the reader needs to run only step 4 in \verb|raman/defect| directory for the defect-induced Raman calculation. It is important to note that the reader needs to copy and paste the files \verb|phonon_sort| (in step 2) from the \verb|phonon| directory, and files \verb|crystal.fmt|, \verb|dmedata.fmt|, \verb|epwdata.fmt|, and \verb|Gra.ukk| (in step 3) from the \verb|epw| directory to the \verb|raman/defect| directory before running step 4.

\noindent\ding{111}\enspace\textbf{Input files:} The input file \verb|epw-raman.in| for step 4 is given as follows:
\begin{codeQEsmall}{examples/graphene/raman/defect/epw-raman.in}
epw
&inputepw
  prefix       = 'Gra'
  outdir       = '../../tmp/'
  dvscf_dir    = '../../epw/save/'
  system_2d    = 'dipole_sp'
  asr_typ      = 'simple'
  vme          = 'dipole'

  nbndsub      = 5
  ep_coupling  = .true.
  elph         = .true.
  use_ws       = .true.
  num_iter     = 400

  fsthick      = 5
  eps_acustic  = 5
  degaussw     = 0.005
  degaussq     = 0.01

  epwwrite     = .false.
  epwread      = .true.
  wannierize   = .false.

  nk1  = 30,   nk2  = 30,   nk3  = 1
  nq1  = 15,   nq2  = 15,   nq3  = 1
  
  nkf1 = 150,  nkf2 = 150,  nkf3 = 1
  nqf1 = 150,  nqf2 = 150,  nqf3 = 1

  Raman_type    = 'defect'
  lhoph         = .true.
  Elaser        = 1.96
  Egamma        = 0.1
  reson_lim     = .true.
  reson_thr     = 1
  prtdipole     = .true.
  prtRaman      = .true.
  polar         = 'all'
  filraman      = 'qraman'
  Cq = 0.d0, 0.d0, 0.d0, 1.d-2, 0.d0, 0.d0, 0.d0
 /
\end{codeQEsmall}
We used the interpolated $\bm k$-points and $\bm q$-points of $150 \times 150 \times 1$ at lines 28 and 29 based on the convergence test of the SRR and DRR calculations. At line 31, we set \verb|Raman_type = 'defect'| for the defect-induced Raman calculation, in which the constant electron-defect coupling is given at line 41. Here, electron-defect coupling, $\mathcal{M}_\text{e-d}$, is expressed by the polynomial function of $q$ is as follows: $\mathcal{M}_\text{e-d}(\bm q)= aq^{-3}+bq^{-2}+cq^{-1}+d+eq+fq^{2}+gq^{3}$, for a future purpose. As discussed in Sec.~\ref{sec:I-defect}, we adopt $\mathcal{M}_\text{e-d}$ as a constant in the present version of {\sc QR$^2$-code}. Thus, only $d$ has a non-zero value (at line 41), which can be obtained by fitting the calculated Raman spectra with existing experimental data.

The input file \verb|raman_pp.in| in the step 4 is given as follows:
\begin{codeQEsmall}{examples/Graphene/raman/defect/raman\_pp.in}
&PLOT
  dir_raman      = './qraman-xx'
  Raman_type     = 'defect'
  lhoph          = .true.
  lRay_sca       = .false.
  Ray_thr        = 5
  Rs_min         = 0
  Rs_max         = 4000
  Rs_inc         = 0.1
  Lgamma         = 20
/
\end{codeQEsmall}
At line 3, we set \verb|Raman_type = 'defect'| for the defect-induced Raman calculation. Other lines are similar to \verb|raman_pp1.in| for the DRR calculation.  

\begin{figure}[t]
  \centering 
  \includegraphics[width=0.9\linewidth]{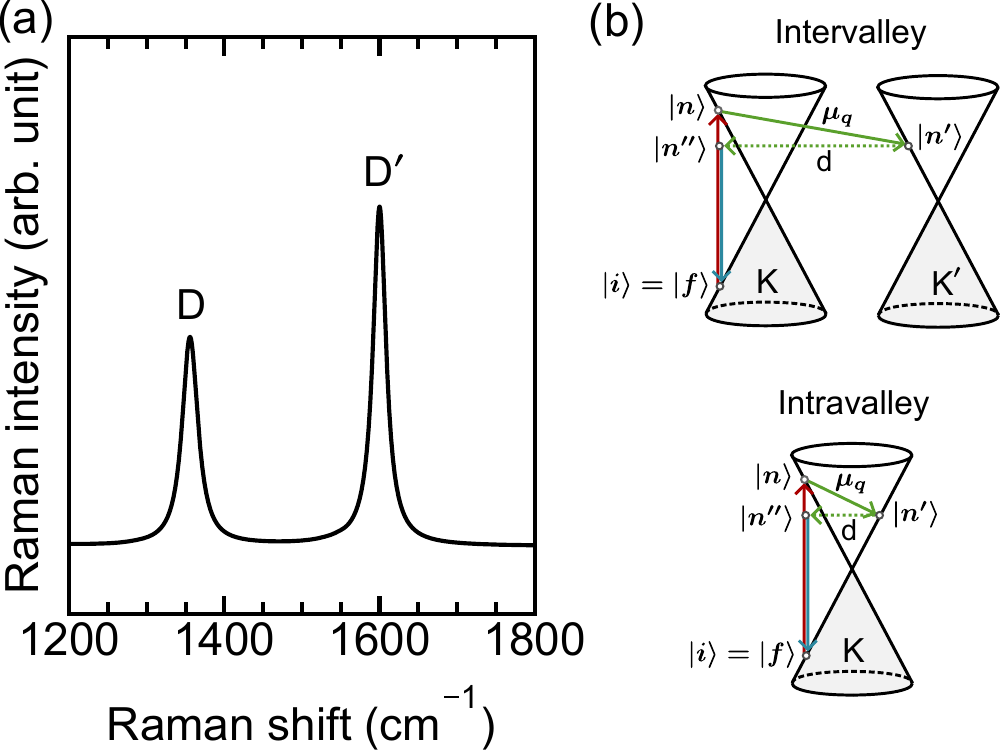}
  \caption{\label{fig:Defect} (a) defect-induced Raman spectra of graphene at 1.96 eV laser energy, in which two peaks correspond to D and D$^\prime$ modes. (b) Schematic of a phonon–defect intervalley and intravalley scattering for the $ee$-process, where the defect scattering is represented by the dotted line.}
\end{figure}

\noindent\ding{111}\enspace\textbf{Output files:}
The Raman spectra in the output file \verb|Raman.dat| are calculated for the laser energy of 1.96 eV, which has a similar format to the output file \verb|Raman.dat| of the DRR calculation in Sec. 4.2. In Fig.~\ref{fig:Defect}(a), we plot the total Raman spectra as a function of the Raman shift at $E_\mathrm{L} = 1.96$ eV. Two bands, D and D$^\prime$, are found at 1355.7 cm$^{-1}$ and 1600.0 cm$^{-1}$, respectively, which are consistent with observed defect-activated peaks in experiment~\cite{eckmann2012probing}. These new Raman peaks are the result of intervalley, intravalley, and a combination of both processes (see Fig.~\ref{fig:Defect}(b)), where the conservation of momentum is preserved by the presence of a defect~\cite{gontijo2019double}.

\section{Summary}
\label{sec:summary}
In summary, we developed a new open-source code {\sc QR$^2$-code} for calculating double-resonance Raman spectra based on the output of {\sc Quantum ESPRESSO} and {\sc EPW}, both of which are free and widely used software. We explain how to download and install the program. Compared with our previous open-source code {\sc QERaman}, which calculates single-resonance Raman spectra, {\sc QR$^2$-code} has demonstrated several advanced capabilities, such as simultaneous calculation of SRR, DRR, defect-induced Raman, and assignment of two-phonon mode calculations. We also provide an example of graphene for using {\sc QR$^2$-code}. Other examples of MoS$_2$ and h-BN monolayers are available on the GitHub page. The program can be used by anybody to analyze the observed resonance Raman spectra of low-dimensional materials. The authors welcome any feedback for improvement from users. 


\section*{Acknowledgement}
T.Y. and J.Q.H. acknowledge the National Natural Science Foundation of China Grants (Nos. 12404213, 52031014), the Strategic Priority Research Program of the Chinese Academy of Sciences (Grant No. XDB0460000), and the National Key R\&D Program of China (2022YFA1203901). N.T.H. acknowledges financial support from the Frontier Research Institute for Interdisciplinary Sciences, Tohoku University. R.S. acknowledges the Yushan Fellow Program by the Ministry of Education (MOE), Taiwan. The simulation work was carried out at National Supercomputer Center in Tianjin, China, and the calculations were performed on TianHe-1(A).

\bibliographystyle{elsarticle-num}

\begin{thebibliography}{10}
\expandafter\ifx\csname url\endcsname\relax
  \def\url#1{\texttt{#1}}\fi
\expandafter\ifx\csname urlprefix\endcsname\relax\def\urlprefix{URL }\fi
\expandafter\ifx\csname href\endcsname\relax
  \def\href#1#2{#2} \def\path#1{#1}\fi

\bibitem{jorio2011raman}
A.~Jorio, M.~S. Dresselhaus, R.~Saito, G.~Dresselhaus, Raman spectroscopy in graphene related systems, John Wiley \& Sons, 2011.

\bibitem{malard2009raman}
L.~M. Malard, M.~A. Pimenta, G.~Dresselhaus, M.~S. Dresselhaus, Raman spectroscopy in graphene, Phys. Rep. 473~(5-6) (2009) 51--87.

\bibitem{i1335}
R.~Saito, M.~Hofmann, G.~Dresselhaus, A.~Jorio, M.~S. Dresselhaus, Raman spectroscopy of graphene and carbon nanotubes, Advances in Physics 60 (2011) 413--550.

\bibitem{hung2024qeraman}
N.~T. Hung, J.~Huang, Y.~Tatsumi, T.~Yang, R.~Saito, {QERaman: An open-source program for calculating resonance Raman spectra based on Quantum ESPRESSO}, Comput. Phys. Commun. 295 (2024) 108967.

\bibitem{giannozzi09-espresso}
P.~Giannozzi, S.~Baroni, N.~Bonini, M.~Calandra, R.~Car, C.~Cavazzoni, D.~Ceresoli, G.~L. Chiarotti, M.~Cococcioni, I.~Dabo, A.~D. Corso, S.~de~Gironcoli, S.~Fabris, G.~Fratesi, R.~Gebauer, U.~Gerstmann, C.~Gougoussis, A.~Kokalj, M.~Lazzeri, L.~Martin-Samos, N.~Marzari, F.~Mauri, R.~Mazzarello, S.~Paolini, A.~Pasquarello, L.~Paulatto, C.~Sbraccia, S.~Scandolo, G.~Sclauzero, A.~P. Seitsonen, A.~Smogunov, P.~Umari, R.~M. Wentzcovitch, {\sc Quamtum ESPRESSO}: A modular and open-source software project for quantum simulations of materials, J. Phys. Condens. Matter 21~(39) (2009) 395502.

\bibitem{huang2022first}
J.~Huang, H.~Guo, L.~Zhou, S.~Zhang, L.~Tong, R.~Saito, T.~Yang, Z.~Zhang, First-principles calculations of double resonance {Raman} spectra for monolayer {MoTe$_2$}, Phys. Rev. B 105 (2022) 235401.

\bibitem{zhang2025duv}
Y.~Zhang, R.~Liu, J.~Huang, N.~Tuan~Hung, R.~Saito, T.~Yang, Z.~Zhang, {DUV} double-resonant {Raman} spectra and interference effect in graphene: First-principles calculations, J. Raman Spectrosc. 56~(4) (2025) 316--323.

\bibitem{saito2024deep}
R.~Saito, N.~T. Hung, T.~Yang, J.~Huang, H.-L. Liu, D.~P. Gulo, S.~Han, L.~Tong, Deep-ultraviolet and helicity-dependent {Raman} spectroscopy for carbon nanotubes and {2D} materials, Small (2024) 2308558.

\bibitem{gontijo2019temperature}
R.~N. Gontijo, A.~Gadelha, O.~J. Silveira, B.~R. Carvalho, R.~W. Nunes, L.~C. Campos, M.~A. Pimenta, A.~Righi, C.~Fantini, Temperature dependence of the double-resonance {Raman} bands in monolayer {MoS$_2$}, J. Raman Spectrosc. 50~(12) (2019) 1867--1874.

\bibitem{liu2024helicity}
R.~Liu, L.-H. Li, Y.~Zhang, J.~Huang, M.-L. Lin, N.~T. Hung, Z.~Wang, Z.~Zhang, R.~Saito, P.-H. Tan, T.~Yang, Helicity selection rule of double resonance {Raman} spectra for monolayer {MoSe$_2$}, Phys. Rev. B 110~(24) (2024) 245422.

\bibitem{thomsen2000double}
C.~Thomsen, S.~Reich, Double resonant {Raman} scattering in graphite, Phys. Rev. Lett. 85~(24) (2000) 5214.

\bibitem{kurti2002double}
J.~K{\"u}rti, V.~Z{\'o}lyomi, A.~Gr{\"u}neis, H.~Kuzmany, Double resonant {Raman} phenomena enhanced by van hove singularities in single-wall carbon nanotubes, Phys. Rev. B 65~(16) (2002) 165433.

\bibitem{saito2003double}
R.~Saito, A.~Gr{\"u}neis, G.~G. Samsonidze, V.~Brar, G.~Dresselhaus, M.~Dresselhaus, A.~Jorio, L.~Can{\c{c}}ado, C.~Fantini, M.~Pimenta, et~al., Double resonance {Raman} spectroscopy of single-wall carbon nanotubes, New J. Phys. 5~(1) (2003) 157.

\bibitem{dresselhaus2005raman}
M.~S. Dresselhaus, G.~Dresselhaus, R.~Saito, A.~Jorio, Raman spectroscopy of carbon nanotubes, Phys. Rep. 409~(2) (2005) 47--99.

\bibitem{jorio2005resonance}
A.~Jorio, C.~Fantini, M.~Pimenta, R.~Capaz, G.~G. Samsonidze, G.~Dresselhaus, M.~Dresselhaus, J.~Jiang, N.~Kobayashi, A.~Gr{\"u}neis, et~al., Resonance {Raman} spectroscopy (n, m)-dependent effects in small-diameter single-wall carbon nanotubes, Phys. Rev. B 71~(7) (2005) 075401.

\bibitem{zhang2022quantum}
S.~Zhang, J.~Huang, Y.~Yu, S.~Wang, T.~Yang, Z.~Zhang, L.~Tong, J.~Zhang, Quantum interference directed chiral {Raman} scattering in two-dimensional enantiomers, Nat. Commun. 13~(1) (2022) 1254.

\bibitem{huang2022new}
J.~Huang, Z.~Liu, T.~Yang, Z.~Zhang, New selection rule of resonant {Raman} scattering in {MoS$_2$} monolayer under circular polarization, J. Mater. Sci. Technol. 102 (2022) 132--136.

\bibitem{pang2022accurate}
Y.~Pang, J.~Huang, T.~Yang, Z.~Zhang, Accurate assignment of double resonant {Raman} bands in {Janus MoSSe} monolayer from first-principles calculations, J. Mater. Sci. Technol. 131 (2022) 82--90.

\bibitem{ponce2016epw}
S.~Ponc{\'e}, E.~R. Margine, C.~Verdi, F.~Giustino, {\sc EPW}: Electron--phonon coupling, transport and superconducting properties using maximally localized wannier functions, Comput. Phys. Commun. 209 (2016) 116--133.

\bibitem{f1020}
R.~Saito, A.~Gr{\"u}neis, G.~G. Samsonidze, V.~W. Brar, G.~Dresselhaus, M.~S. Dresselhaus, A.~Jorio, L.~G. Can\c{c}ado, C.~Fantini, M.~A. Pimenta, A.~G. {Souza~Filho}, Double resonance {Raman} spectroscopy of single wall carbon nanotubes, New J. Phys. 5 (2003) 157.1--157.15.

\bibitem{i1179}
M.~A. Pimenta, G.~Dresselhaus, M.~S. Dresselhaus, L.~G. Can\c{c}ado, A.~Jorio, R.~Saito, Studying disorder in graphite-based systems by {Raman} spectroscopy, Phys. Chem. Chem. Phys. 9 (2007) 1276--1291.

\bibitem{n950}
A.~Gr\"uneis, R.~Saito, G.~G. Samsonidze, T.~Kimura, M.~A. Pimenta, A.~Jorio, A.~G.~S. Filho, G.~Dresselhaus, M.~S. Dresselhaus, Inhomogeneous optical absorption around the {K} point in graphite and carbon nanotubes, Phys. Rev. B 67 (2003) 165402.

\bibitem{hung2022quantum}
N.~T. Hung, A.~R. Nugraha, R.~Saito, Quantum ESPRESSO Course for Solid-State Physics, Jenny Stanford Publishing, 2022.

\bibitem{ferrari2006raman}
A.~C. Ferrari, J.~C. Meyer, V.~Scardaci, C.~Casiraghi, M.~Lazzeri, F.~Mauri, S.~Piscanec, D.~Jiang, K.~S. Novoselov, S.~Roth, et~al., Raman spectrum of graphene and graphene layers, Phys. Rev. Lett. 97~(18) (2006) 187401.

\bibitem{liu2015deep}
H.-L. Liu, S.~Siregar, E.~H. Hasdeo, Y.~Kumamoto, C.-C. Shen, C.-C. Cheng, L.-J. Li, R.~Saito, S.~Kawata, {Deep-ultraviolet Raman scattering studies of monolayer graphene thin films}, Carbon 81 (2015) 807--813.

\bibitem{cong2018stokes}
X.~Cong, J.-B. Wu, M.-L. Lin, X.-L. Liu, W.~Shi, P.~Venezuela, P.-H. Tan, {Stokes and anti-Stokes Raman scattering in mono-and bilayer graphene}, Nanoscale 10~(34) (2018) 16138--16144.

\bibitem{venezuela2011theory}
P.~Venezuela, M.~Lazzeri, F.~Mauri, {Theory of double-resonant Raman spectra in graphene: Intensity and line shape of defect-induced and two-phonon bands}, Phys. Rev. B 84~(3) (2011) 035433.

\bibitem{eckmann2012probing}
A.~Eckmann, A.~Felten, A.~Mishchenko, L.~Britnell, R.~Krupke, K.~S. Novoselov, C.~Casiraghi, Probing the nature of defects in graphene by raman spectroscopy, Nano Lett. 12~(8) (2012) 3925--3930.

\bibitem{gontijo2019double}
R.~N. Gontijo, G.~C. Resende, C.~Fantini, B.~R. Carvalho, Double resonance raman scattering process in 2d materials, J. Mater. Res. 34~(12) (2019) 1976--1992.

\end{thebibliography}

\end{document}